\documentclass{aa}  

\usepackage{graphicx}
\usepackage{txfonts,textcomp}
\usepackage{placeins}           

\begin{document}

   \title{First detection of HDO ice in a protoplanetary disk}

%

 \author{Alexey Potapov\inst{1}\fnmsep\thanks{Corresponding author: alexey.potapov@uni-jena.de}
        \and Piyush Kalambkar\inst{1}
        \and Jeroen Bouwman\inst{2}
        \and Christiaan Boersma\inst{3}
        \and Hiroshi Terada\inst{4}
        \and Will R. Rocha\inst{5}
        \and Hendrik Linz\inst{2}
        }

   \institute{Analytical Mineralogy Group, Institute of Geosciences, Friedrich Schiller University Jena, Jena, Germany
   \and Max Planck Institute for Astronomy, Heidelberg, Germany
   \and NASA Ames Research Center, USA
    \and National Astronomical Observatory of Japan, National Institutes of Natural Sciences (NINS), Tokyo, Japan
    \and Laboratory for Astrophysics, Leiden Observatory, Leiden University, The Netherlands}

   \date{Received February 13, 2026, Accepted June 03, 2026}

 
  \abstract
   {Protoplanetary disks are the birthplace of planets and planetary systems. Investigating the molecular inventory of disks is key to linking the chemical evolution of the interstellar medium and the makeup of planets and their atmospheres. In particular, tracing the history of the deuterium enrichment of water along the journey from interstellar clouds through protoplanetary disks to planetary systems provides critical insights into the chemical inheritance.}
   {We aim to investigate the chemical composition of ices in protoplanetary disks; specifically, the presence of HDO ice that ought to be present, but has not been detected in disks thus far.} 
   {We analyzed JWST/NIRSpec observations of the 132-1832 edge-on disk located in the Orion Nebula Cluster using the ENIIGMA fitting tool and unique laboratory data.}
   {We report on the first detections of HDO ice in a protoplanetary disk. The estimated upper limit for the HDO/H\textsubscript{2}O ratio for 132-1832 is much higher, compared to HDO/H\textsubscript{2}O ratios obtained for chondrites, comets, and embedded young stellar objects. In the disk ices, beyond HDO, we detected H\textsubscript{2}O, CO\textsubscript{2}, $^1$$^3$CO\textsubscript{2}, CO, OCN$^-$, and OCS, species, whose presence has also been detected in other disks.}
   {The HDO ice detection may point to the efficient ice processing in the disk and confirm the findings of laboratory experiments on deuterated ices.}


   \maketitle
   \nolinenumbers

\section{Introduction}
The outstanding spectral sensitivity, broad wavelength coverage, and high spatial resolution of the James Webb Space Telescope (JWST) is enabling significant progress in exploring the chemical compositions of various astrophysical environments, including protoplanetary disks (see, e.g., \citealt{Henning2024}). The JWST/NIRSpec observations of the edge-on disk HH 48 NE revealed spatially resolved absorption features of the major ice components H\textsubscript{2}O, CO\textsubscript{2}, and CO, multiple weaker signatures from the less abundant ices NH\textsubscript{3}, OCN$^-$, and OCS. In addition, it has offered, for the first time in a protoplanetary disk, a clear detection of $^{13}$CO\textsubscript{2} ice \citep{RN1738}. JWST/MIRI observations of the same disk provided evidence for the presence of CH\textsubscript{4} ice and an upper limit for CH\textsubscript{3}OH ice \citep{RN1777}. Using the same instrument a tentative detection of CO\textsubscript{2} ice was made toward a tiny brown dwarf disk \citep{Perotti2025}. The JWST NIRSpec+MIRI observations of the edge-on disk d216-0939 \citep{RN1876} reported detections of H\textsubscript{2}O, CO\textsubscript{2}, $^{13}$CO\textsubscript{2}, CO, OCN$^-$, and MgSiO\textsubscript{3}/Mg\textsubscript{2}SiO\textsubscript{4} amorphous silicates and, tentatively, $^{13}$CO and OCS, as well as, for the first time in protoplanetary disks, NH$_4^+$ and the complex organic molecule ammonium carbamate (NH$_4^+$NH\textsubscript{2}COO$^-$). The latter detection stands as a direct confirmation of increased production of complex organic molecules (COMs) in warmer astrophysical environments, suggested by laboratory experiments and models \citep{RN983,RN66}. Also, H\textsubscript{2}O, CO\textsubscript{2}, and CO ice absorption features were detected in the spectra of the edge-on disks of Tau 042021 \citep{RN2020}, FS Tau B, HH 30, IRAS 04302, and Tau 042021 \citep{RN2021}. From the imaging point of view, water ice was detected in the d114–426 disk with JWST/NIRCam \citep{Ballering2025}. 

Characterizing the deuterium enrichment of water, quantified via the HDO/H\textsubscript{2}O ratio in young stellar objects (YSOs) and their protoplanetary disks, offers key insights into whether the material that forms planets is inherited from the interstellar medium (ISM) in a pristine form or reprocessed during disk evolution. 
On the envelope scale, submillimeter observations toward the low-mass protostar IRAS~16293–2422 revealed HDO gas in both ground-state and excited transitions, with an abundance ratio in the warm inner envelope of a few $\times 10^{-4}$, comparable to that observed in comets (e.g., \citealt{Eberhardt1995, Altwegg2015, 2025NatAs.tmp..159C}). This suggests that prestellar deuteration might be largely preserved through protostellar collapse and disk formation \citep{Stark2004}.
Interferometric studies using the Plateau de Bure and the SubMillimeter Array (SMA) toward low-mass protostars such as NGC1333-IRAS2A and IRAS4B have measured warm gas-phase HDO/H\textsubscript{2}O ratios around $10^{-3}$--$6\times10^{-4}$, consistent with cometary and oceanic values, albeit with some indications of in situ gas-phase processing in the regions making up the inner envelope \citep{Coutens2014, Jorgensen2010}.
More recently, observations with the Herschel Space Telescope and especially with the Atacama Large (sub-)Millimeter Array (ALMA) reinforced the idea of inheritance. They showed that the HDO/H\textsubscript{2}O and D\textsubscript{2}O/HDO ratios remain similar to prestellar values across a range of environments, supporting a dominant inheritance scenario over complete in-disk reprocessing \citep{Furuya2016, vanDishoeck2021}. These higher angular-resolution observations also showed that the ratios measured toward isolated sources \citep{Jensen2019, Jensen2021b, Andreu2023} are higher by factors of 2–10 when compared to clustered protostars \citep{Persson2014}. This reveals the influence the environment has on the initial deuteration.

The latest breakthroughs with JWST have enabled the first detection of the HDO ice feature at 4.1~\textmu m toward both massive and isolated low-mass protostars. \citeauthor{Slavicinska2024} measured HDO/H\textsubscript{2}O ice ratios of $(4.4^{+3.7}_{-1.7})\times10^{-3}$ toward L1527 (isolated) and similarly toward massive protostars. This indicates that ice-phase deuteration levels are largely inherited into protostellar envelope material with limited alteration \citep{Slavicinska2024, Slavicinska2025}. 

On the disk scale, the first detection of HDO gas was reported in \citeyear{Ceccarelli2005} toward the Class~II disk DM Tau by \citeauthor{Ceccarelli2005}, demonstrating unexpected deuterated water vapor at radii beyond theoretical ice lines. This suggests that deuterium fractionation may undergo a reset during the disk phase \citep{Ceccarelli2005}. Since then, deuterated water vapour has only been detected toward one, younger disk, namely the outbursting Class~I V883~Ori disk \citep{Tobin2023}.

Due to the poor sensitivity and high systematic error in ground-based M-band, coupled with the lack of spatial resolution of pre-JWST space observations, HDO-ice remained undetected in Class~II disks, a critical stage linking embedded YSOs (Class~0 and I) to fully formed planetary systems. A tentative assignment was reported toward the triplet system HV~Tau using the AKARI telescope \citep{Aikawa2012}. Since the full width at half maximum (FWHM) of the AKARI point spread function (PSF) is 4.\textquotedbl 3 and the edge-on disk HV Tau C is separated from the binary (HV Tau A and B) by approximately 4\textquotedbl, it was not possible to disentangle whether the HDO absorption feature originated exclusively from the HV Tau C disk. 

Against this background, the Class II protoplanetary disk 132-1832 in the Orion Nebula Cluster (ONC; $d=390{\pm}2~$pc; \citealt{MaizApellaniz_2022}) represents a particularly favorable target for disk ice studies. Initially identified as a pure silhouette disk without any nebulosities by \citet{2000AJ....119.2919B}, the disk has a high inclination of 75$^\circ$. The properties of the disk are summarized in Table~\ref{tab:disk_properties} in the Appendix. The inclined configuration was confirmed by \citet{2012AJ....144..175T}, which also reported the detection of the 3-\textmu m water ice absorption band in the disk, while showing no sign of water ice toward a nearby (5\textquotedbl7) star. The high inclination makes this disk well suited for JWST observations of the main ice components and their isotopologs. Here, we report on the first ever detection of HDO ice in the protoplanetary disk 132-1832 with the JWST/NIRSpec instrument and other ice species detections.

\section{Methodology}
\subsection{Observations and analysis}
The highly inclined protoplanetary disk 132-1832 in the Orion Nebula Cluster was observed within the Cycle 1 General Observation (GO) program 1741 (PI: A. Potapov). For details of the observations and data reduction we refer to \citet{RN1876}, as identical data reduction and analysis procedures are used as described there. In brief, observations were carried out using the integral field units (IFU) of the NIRSpec instrument using the medium-resolution gratings (G235M/F170LP and G395M/F290LP) and a four-point dither strategy, to cover the 2–5~\textmu m wavelength range. Target acquisition (TA) on nearby GAIA-referenced stars were used to ensure the precise positioning required because of the source’s extended nature at shorter wavelengths. LeakCal observations were implemented for the NIRSPEC IFU observations to correct for possible contamination from bright nearby emission sources.

Data reduction was performed using the JWST pipeline \texttt{v1.13.4} \citep{Bushouse2024}, following the pmap\_1210 calibration. Spectral extraction was done via the \texttt{extract1D} step using a fixed width of 2$\times$FWHM for the extraction aperture. In the resulting 1D spectra, a broad absorption feature is visible at the spectral location of the PAH band at around 3.29~\textmu m \citep{2008ARA&A..46..289T}. The occurrence of this feature is not an artifact of an incorrect background subtraction. It is a real feature independent of the chosen background spectrum, as we show in Figure~\ref{fig:background} in the Appendix. The highly inclined circumstellar disk absorbs part of the extended PAH background emission from the Orion environment along its major axis. This leaves an absorption imprint in the background-subtracted source spectrum. While a very interesting effect in itself, it is beyond the scope of this paper. For our final spectrum, we relied on the classic implemented background evaluation method with a larger annulus around the source aperture, which partly accounts for the large-scale emission gradient. The inner and outer radii of the annulus are 1.0 and 1.2 arcsec, respectively.

To analyze the observed ice absorption bands, a continuum estimation was made using a third-order polynomial for the NIRSpec wavelength range, with continuum points of 2.5 - 2.55, 2.6 - 2.65, 2.69 - 2.70, 4.0 - 4.04, 4.06 - 4.07, 4.6 - 5.0, and 5.15 - 5.2~\textmu m. We note that we explicitly neglected the footpoints close to the PAH features around 3.2--3.4~\textmu m. Variations in the polynomial order and continuum anchor points were considered, but they were not shown to exert any influence on the band assignments presented here.

Spectral decomposition of ice features was performed using the \texttt{ENIIGMA} tool \citep{RN1441} that fits optical depth spectra by scaling laboratory absorbance data. The goodness-of-fit was quantified using both the root-mean-square-error (RMSE) and the Akaike information criterion (AIC). This allowed us to compare different models, even when they have a varying numbers of free parameters. The criteria for the best fit were: (i) a decrease in the RMSE and (ii) a decrease in the AIC or an increase by no more than 2 with respect to the lowest value \citep{Burnham2002}. 

The scattering spectra were calculated using the \texttt{OpTool}
computational code \citep{Dominik2021}. For the core component, we adopted optical constants of MgSiO\textsubscript{3} grains mixed with H\textsubscript{2}O ice at 10~K from laboratory measurements \citep{Potapov2018}. For the mantle component, we considered pure CO\textsubscript{2} ice and H\textsubscript{2}O:CO\textsubscript{2} (10:1) ice, modeled using laboratory derived optical constants from \citep{Warren1986} and \citet{2017MNRAS.464..754R} correspondingly. We applied the DHS (distribution of hollow spheres) grain shape approach. The models were computed using grain size distributions spanning $a_{\min}=0.005~\mu$m up to $a_{\max}=1$--$5~\mu$m with power law indices $-3.5$ and $-2.0$.

Once the best fits were defined, we calculated the column density using ENIIGMA for each component as the band area divided by the vibrational mode band strength of the molecule following the procedure described in \citep{RN1917}. As noted there, a problem of the band strength determination is its dependence on the ice density, which gives typical uncertainties of around 15 and 30\% for pure and mixed ices, respectively. For the determination of the H\textsubscript{2}O and HDO column densities, we used the band strengths of 2.1$\times$10$^{-16}$~cm/molecule for 50~K water ice from \citep{RN96} and 4.1$\times$10$^{-17}$~cm/molecule for 14~K HDO ice from \citep{Galvez2011}. There is also a contribution of trapped water \citep{RN1715} to the H\textsubscript{2}O column density. The band strength of trapped water is not known and we used the value for 10~K water ice, 1.9$\times$10$^{-16}$~cm/molecule \citep{RN96}. Taking into account the uncertainties of the band strengths and continuum determinations, we considered a column density uncertainty of 50\% for the presented H\textsubscript{2}O and HDO values.  

\subsection{Laboratory measurements}
The infrared spectra of three-component mixtures, MgSiO\textsubscript{3}/H\textsubscript{2}O/NH\textsubscript{3} and MgSiO\textsubscript{3}/H\textsubscript{2}O/CH\textsubscript{3}OH, at various temperatures were specially measured for the present study (the spectra will be published separately). The mixtures were produced in the Jena Dust Machine, allowing for a simultaneous deposition of nanoparticles (analogs of cosmic dust grains) and molecular ices. Detailed descriptions of the technique and dust samples have been presented in previous publications \citep[e.g.,][and references therein]{Potapov2018, RN1974}. In brief, the formation of nm-sized MgSiO\textsubscript{3} particles was performed by pulsed laser ablation of a Mg:Si target and the subsequent condensation of evaporated species in a quenching atmosphere of O\textsubscript{2}. The condensed particles were extracted adiabatically from the ablation chamber, generating a particle beam that was directed into a separate chamber, where silicate grains were deposited onto a substrate, forming highly porous fractal aggregates. Simultaneously, H\textsubscript{2}O/NH\textsubscript{3} or H\textsubscript{2}O/CH\textsubscript{3}OH ices premixed in the ratios of 10:1 and 12:1 correspondingly were deposited. The mixtures were deposited at 10~K and measured at 10, 50, 100 and 150~K after subsequent heating using a Fourier transform infrared (FTIR) spectrometer (Vertex 80v, Bruker) in transmission mode.

\section{Results}
\begin{figure}[ht]
\centering
\includegraphics[width=\columnwidth]{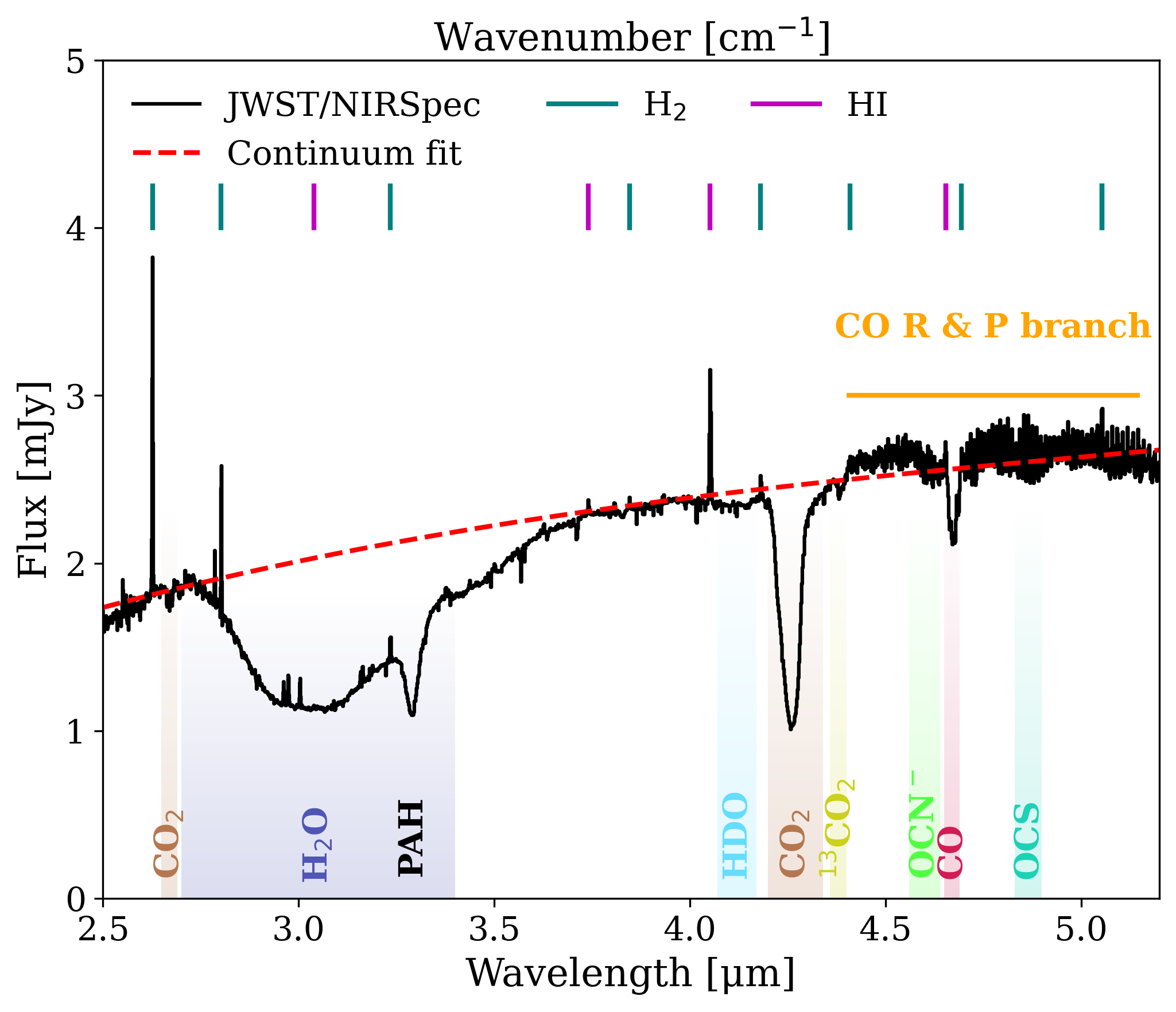}
\caption{JWST/NIRSpec spectrum of the d132-1832 disk. The solid-state bands are highlighted with shaded areas. The major gas lines are also labeled. Zoom-ins into the H$_2$O and HDO ice bands are provided in Figure~\ref{fig:h20bestfit} and~\ref{fig:hdofit}. The continuum is indicated by the dashed red line.}
\label{fig:flux_spectrum}
\end{figure}

\begin{figure}[ht]
\centering
\includegraphics[width=\columnwidth]{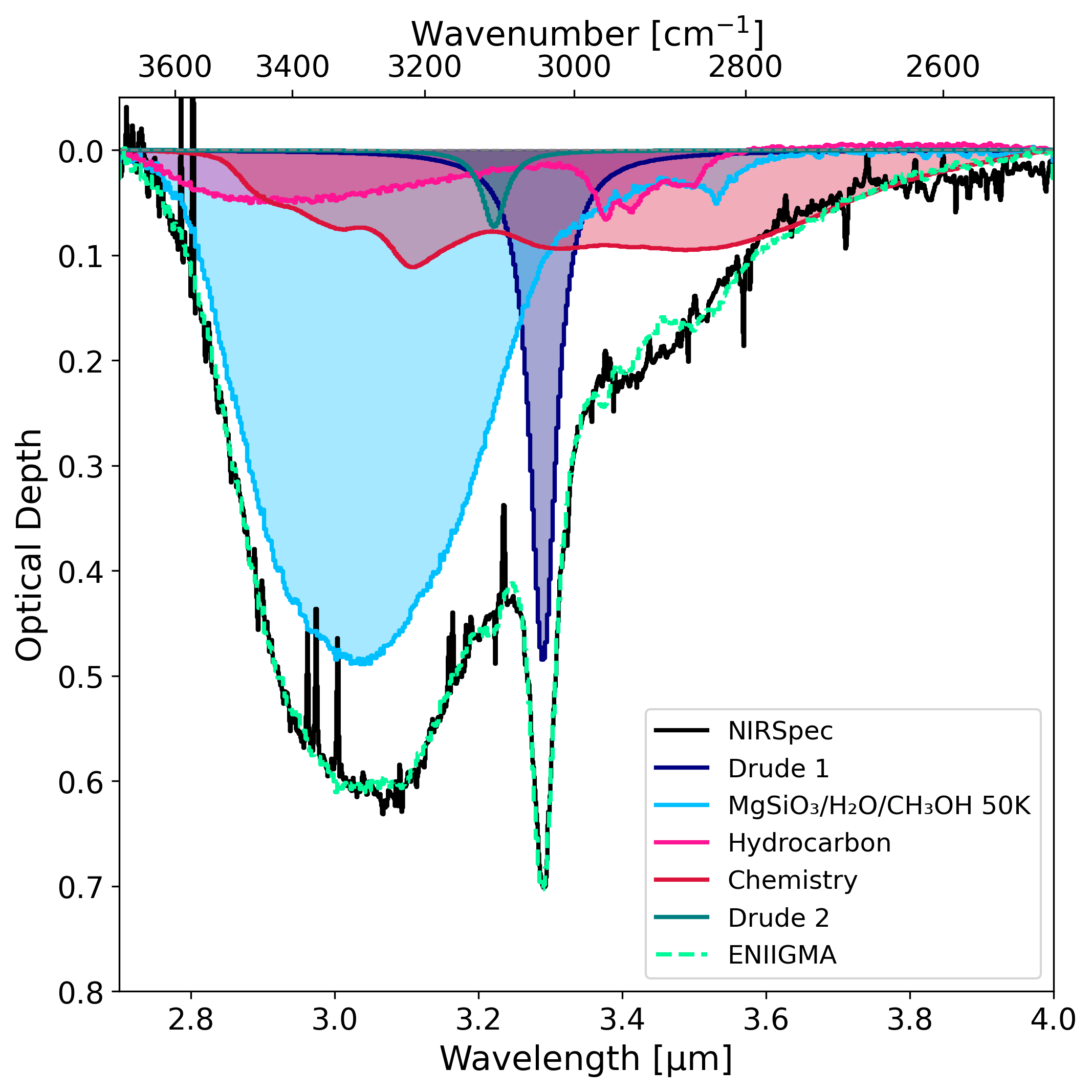}
\caption{Optical depth spectrum and ENIIGMA best-fit model for the 2.7--4.0~\textmu m region and its separate components.}

\label{fig:h20bestfit}
\end{figure}

\begin{figure}[ht!]
\centering
\includegraphics[width=\columnwidth]{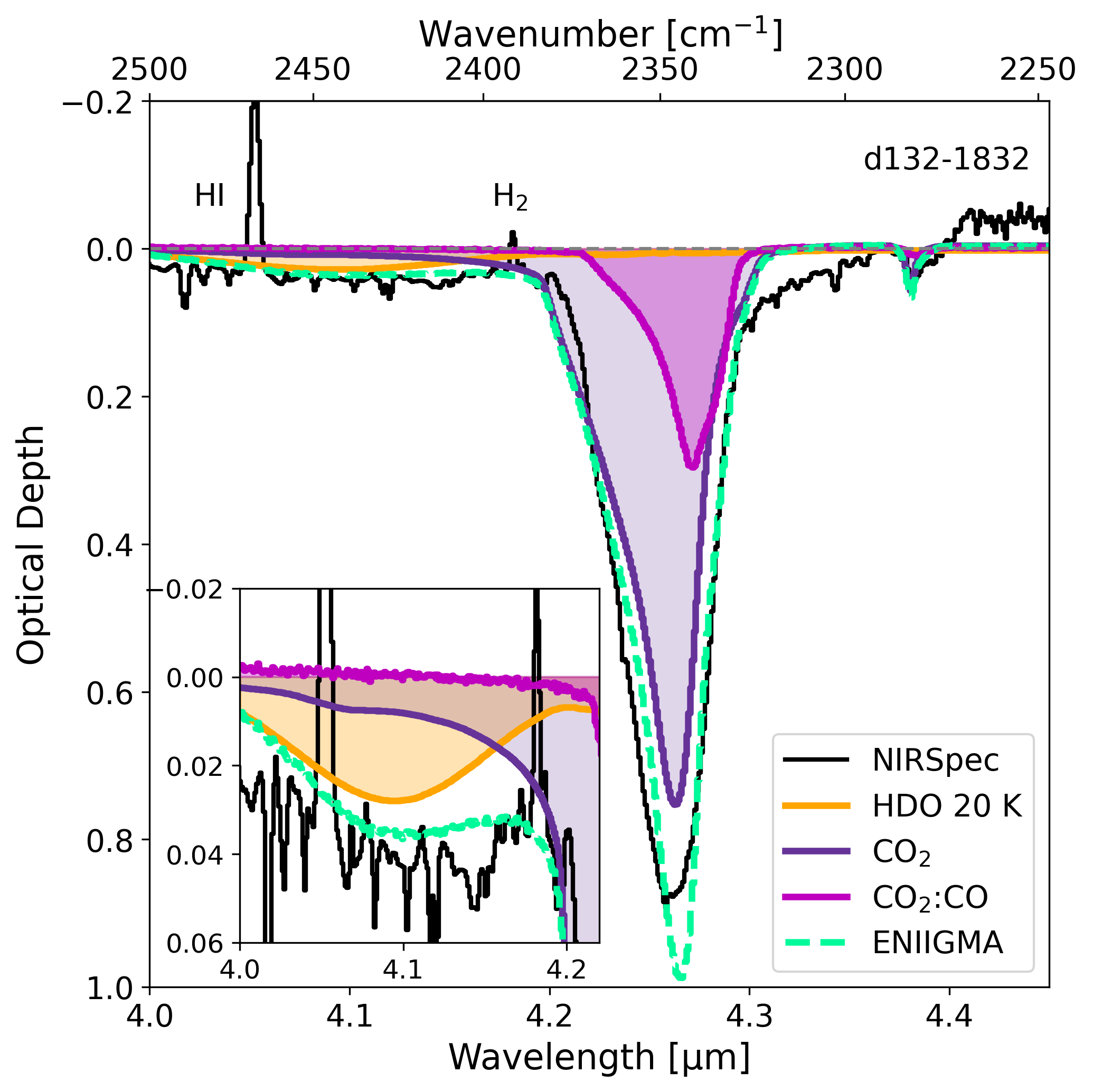}
\caption{Optical depth spectrum and ENIIGMA best-fit model for the 4.0-4.45~\textmu m region and its separate components. Major gas lines are labeled. The inset zooms in on the HDO ice feature.}
\label{fig:hdofit}
\end{figure}

In Figure ~\ref{fig:flux_spectrum}, we present the JWST/NIRSpec spectrum of the 132-1832 disk and assignments of the visible solid-state spectral bands. In the spectrum, the combination mode of CO\textsubscript{2} at 2.68~\textmu m, the H\textsubscript{2}O stretching mode around 3~\textmu m, a PAHs absorption band at 3.29~\textmu m, the HDO mode (O-D stretching) around 4.1~\textmu m, the CO\textsubscript{2} and $^1$$^3$CO\textsubscript{2} stretching modes at 4.26~\textmu m and 4.38~\textmu m, and the spectral signatures of OCN$^-$ at 4.62~\textmu m and CO at 4.67~\textmu m are clearly observed. We also tentatively detected OCS at 4.9~\textmu m. Two regions, 2.7 - 4~\textmu m (H\textsubscript{2}O) and 4.0 - 4.45~\textmu m (HDO+CO\textsubscript{2}), were fitted using the ENIIGMA fitting tool. The best-fit models are shown in Figures~\ref{fig:h20bestfit} and~\ref{fig:hdofit}. The AIC and RMSE values obtained for various fits are presented in Tables~\ref{tab:eniigmafit1} and~\ref{tab:eniigmafit2} in the Appendix. 

\subsection{2.7-4.0 \textmu m region}
\label{subsec:2_7-4_0}
The main spectral band in this region is the H\textsubscript{2}O stretching mode. It might also contain signatures of NH\textsubscript{3}, CH\textsubscript{3}OH, water trapped on silicates \citep{RN873}, and COMs, in particular, ammonium salts such as ammonium carbamate (NH\textsubscript{4}$^+$NH\textsubscript{2}COO$^-$), which had previously been detected by \citet{RN1876}. To model the PAH feature at 3.29~\textmu m, two Drude profiles were used, one centered at 3.29 and the other at 3.22~\textmu m. Their FWHM values are 40 and 38 cm$^{\rm -1}$ for the 3.29 and 3.22~\textmu m component, respectively, which were found through manual iteration. Drude profiles are commonly used to describe PAH (emission) profiles \citep[e.g.,][]{2007ApJ...656..770S, 2021ApJ...917....3D, 2025A&A...701A.111V}.

The AIC and RMSE values obtained for various fits are presented in Table~\ref{tab:eniigmafit1} in the Appendix. Following our previous studies \citep{RN873, RN1876}, we based our analysis of the 2.7-4.0~\textmu m region on the laboratory data obtained for physical mixtures of silicates and ices. We started with combinations of Drude profiles and spectra of MgSiO\textsubscript{3}/H\textsubscript{2}O binary mixtures at different temperatures (10-150~K) that were originally presented in \citet{Potapov2018}. These inputs produced only moderate quality fits, with RMSE values above 0.078 and AIC values above 36. Therefore, they were deemed insufficient to reproduce the observed band.  

Usage of spectra of three-component mixtures, MgSiO\textsubscript{3}/H\textsubscript{2}O/NH\textsubscript{3} and MgSiO\textsubscript{3}/H\textsubscript{2}O/CH\textsubscript{3}OH, instead of spectra of MgSiO\textsubscript{3}/H\textsubscript{2}O mixtures, improved the fit. They lowered the RMSE to $\sim$0.060 and the AIC to $\sim$23--25. The main reason for the improvement is a broadening of the band. However, these combinations alone as well as combinations of spectra taken at different temperatures could not reproduce the red shoulder of the 3~\textmu m feature. A significant improvement was obtained when the chemistry spectrum was included. ``Chemistry'' is the laboratory spectrum obtained after UV irradiation of a mixture of CO\textsubscript{2}+NH\textsubscript{3} at 75~K and its subsequent heating to 230~K \citep{RN983}. The resulting mixture mainly contained ammonium carbamate and carbamic acid and was used for fitting the 3~\textmu m band in our earlier study \citep{RN1876}. 

The best fit from all tried was achieved when hydrocarbons were added on top of the Drude~1 + Drude~2 + MgSiO\textsubscript{3}/H\textsubscript{2}O/CH\textsubscript{3}OH 50~K + chemistry input set. ``Hydrocarbons'' refers to the laboratory spectrum obtained after UV irradiation of a H\textsubscript{2}O+CH\textsubscript{3}OH+MgSiO\textsubscript{3} sample at 10~K and its subsequent heating to 200~K, used for fitting the 3~\textmu m band in \citep{RN873}. The spectrum contains also a trapped water signature. This model yielded RMSE=0.022 and AIC=12.7, and provided the best overall reproduction of both the band depth and the extended red wing of the 3~\textmu m feature. Although the AIC was slightly higher than the chemistry-only case (AIC=11.7), ENIIGMA favored the inclusion of hydrocarbons for the final solution. We therefore adopted this model as the best fit for the 2.7--4.0~\textmu m region, which is shown in Figure~\ref{fig:h20bestfit}.  

\subsection{4.0-4.45 \textmu m region}
The 4.0--4.45~\textmu m region contains the HDO O--D stretch near 4.10~\textmu m, the CO\textsubscript{2} asymmetric stretch at 4.27~\textmu m, and the $^{13}$CO\textsubscript{2} stretching mode at 4.39~\textmu m. The AIC and RMSE values obtained for various fits are presented in Table~\ref{tab:eniigmafit2} in the Appendix. We started our analysis with two inputs, a CO\textsubscript{2} spectrum at 13~K taken from the LIDA database \citep{Rocha2022LIDA} originally presented in \citet{Rocha2014}, and HDO/H\textsubscript{2}O spectra at 14, 90 and 150~K for 0.8\%, 4\%, and 20\% HDO in H\textsubscript{2}O presented in \citet{Galvez2011}. The resulting fits showed similar quality (RMSE=0.055–0.057, AIC=19.4–20.2). By contrast, the combinations of H\textsubscript{2}O/CO\textsubscript{2} and HDO/H\textsubscript{2}O spectra yielded substantially worse fits (RMSE$\geq$0.114, AIC$\geq$ 68). 

In the next step, we extended the models using the CO\textsubscript{2} and HDO/H\textsubscript{2}O spectra with additional components. The inclusion of either H\textsubscript{2}O/CO\textsubscript{2} or CO\textsubscript{2} scattering spectra moderately improved the fit, lowering the AIC to 18.3 and 18.0, respectively. However, these combinations still did not fully reproduce the observed features. A significant improvement was obtained when a spectrum of a CO\textsubscript{2}/CO mixture at 15~K from the LIDA database was added. The models using CO\textsubscript{2}, HDO/H\textsubscript{2}O and CO\textsubscript{2}/CO spectra achieved RMSE=0.046 with AIC between 16.5 and 16.9. The model using the spectra of CO\textsubscript{2}, 20\% HDO in H\textsubscript{2}O at 14~K and CO\textsubscript{2}/CO effectively reproduces both the depth and the extended red wing of the 4.0--4.45~\textmu m feature. This model is shown in Figure~\ref{fig:hdofit}. 

Scattering on icy dust grains can dramatically change the shape of absorption bands. In the CO$_2$ range, scattering appears as a positive bump on the blue side of the CO$_2$ absorption band, as shown in \citep{Dartois2022, RN1876}. Thus, it is close to the HDO band. To test whether the 4.10~\textmu m absorption feature could be affected by CO\textsubscript{2} scattering, we performed a set of OpTool calculations exploring CO\textsubscript{2} scattering scenarios, and replaced in our best-fit model the HDO/H\textsubscript{2}O data by the CO\textsubscript{2} scattering data. In all tested configurations, the CO\textsubscript{2} scattering feature is outside of the HDO absorption region and cannot affect the HDO absorption band profile. The fit models for the edge cases ($a_{\max}=1$ and $5~\mu$m and the power law indices of $-3.5$ and $-2.0$) are shown in Figure~\ref{fig:co2_scattering_AF} in the Appendix. A summary of the CO\textsubscript{2} scattering models is presented in Table~\ref{tab:scattering} in the Appendix.

The addition of further components, such as H\textsubscript{2}O/CO\textsubscript{2}, CO\textsubscript{2} scattering, CH\textsubscript{3}OH 15K from LIDA, and additional HDO/H\textsubscript{2}O temperature component, on top of the best-fit model did not provide any further improvement and only increased the AIC. Next, we applied the HDO/H\textsubscript{2}O data used for fitting the HDO band in \citep{Slavicinska2025}. As can be seen from Table~\ref{tab:eniigmafit2}, the new HDO/H\textsubscript{2}O inputs, HDO(c) and HDO(a), namely, the spectra of amorphous and crystalline ice at various temperatures, did not lead to any improvement of the fit.

\section{Discussion}
\begin{figure*}[ht]
\sidecaption
\includegraphics[width=12cm]{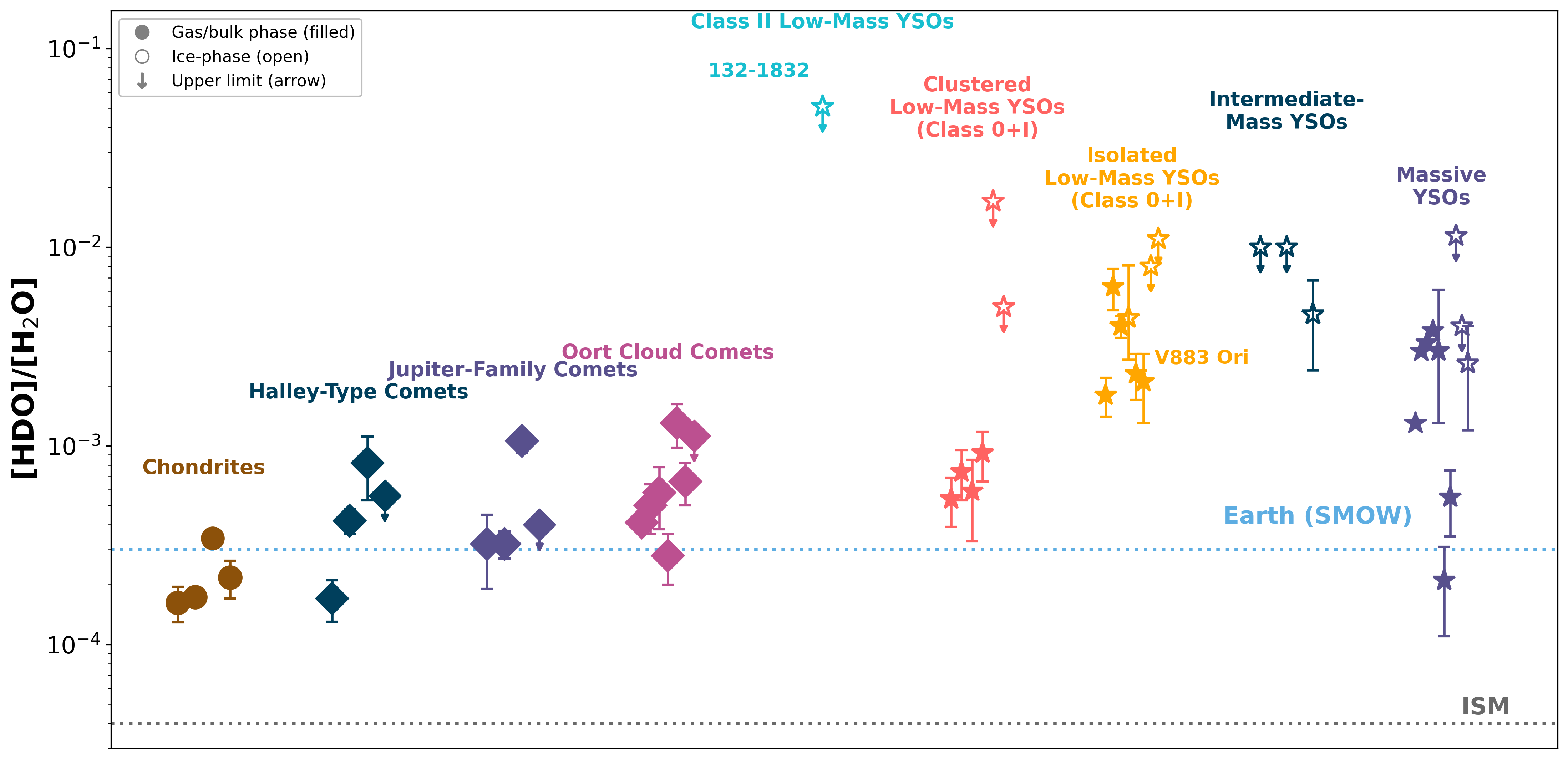}
\caption{Comparison of [HDO]/[H\textsubscript{2}O] ratios toward Solar System objects and protostars. Filled markers represent gas-phase and bulk Solar System values, whereas empty markers show ice measurements. Downward arrows indicate upper limits. For upper-limit ranges, the conservative upper bound is plotted. All values can be found in Table~\ref{tab:C3} and \ref{tab:C4}. The ratio obtained for d132-1832 is labeled.}
\label{fig:ratio}
\end{figure*}

To our knowledge, this study presents the first detection of HDO ice in protoplanetary disks. This result opens new avenues for future search for HDO ice in disks and for a better understanding of the astrochemical links between the ISM, the embedded stages of star formation and Solar System objects. Figure~\ref{fig:ratio} offers a comparison of HDO/H\textsubscript{2}O ratios in both the gas and the solid-state phase toward a plethora of Solar System objects and protostars. 

The H\textsubscript{2}O column density for 132-1832 determined from our best fit is 1.1$\times$10$^{18}$~cm$^{\rm -2}$ and is similar to the column density of 1.15$\times$10$^{18}$~cm$^{\rm -2}$ presented for this disk in a previous observational study with the Subaru telescope \citep{2012AJ....144..175T}. Interestingly, this value is not far from the measurement of 1.9$\times$10$^{18}$~cm$^{\rm -2}$ estimated for the HH48 NE disk \citep{RN1738}, despite the two disks being located in two different star-forming regions (i.e., 132-1832 in the Orion Nebular Cluster and HH48 NE in Chameleon I). The HDO column density for 132-1832 determined from our best fit is 5.6$\times$10$^{16}$~cm$^{\rm -2}$. The resulting HDO/H\textsubscript{2}O ratio of d132-1832 is 5.1$\times$10$^{-2}$, much higher than any other previously determined value. 

We have to note here that the 3~\textmu m H\textsubscript{2}O ice band can be easily saturated in highly inclined disks \citep{RN1740} and its apparent optical depth could thus end up underpredicting the true H\textsubscript{2}O column density. Stellar light can be scattered more through the disk surface rather than through the cold midplane. In such a case, a direct conversion from optical depth to column density is not appropriate for inclined disks and the determined HDO/H\textsubscript{2}O ratio is therefore an upper limit. 

Radiative transfer modeling can be seen as an alternative instruments for ice band detection and column density determination (e.g., \citep{RN1740, RN2047}). Although radiative transfer is a powerful methodology for studying absorption and emission features in protoplanetary disks, it also comes with huge degeneracies when the source properties such as the geometry, density profile, and grain features are not known. In addition, a reliable model would require combining multi-wavelength data from other telescopes. This task is not the main goal of this manuscript and will be saved for future study.

However, in our case, the HDO/H\textsubscript{2}O ratio estimation can be considered reasonable because: (i) there is no evidence of scattering on CO\textsubscript{2}-ice; (ii) there is no evidence of scattering on H\textsubscript{2}O ice; and (iii) the temperature of the H\textsubscript{2}O ice is relatively low. The MgSiO\textsubscript{3}/H\textsubscript{2}O/CH\textsubscript{3}OH 50~K spectrum was used in the best-fit model with the fit tending to lower (10~K), rather than to higher (100 and 150~K) temperatures. Thus, the main H\textsubscript{2}O absorption may take place in the midplane. In case of strong scattering, light mainly passes through warm layers of the disk. For instance, H\textsubscript{2}O ice was fitted using 100 and 150~K spectra in \citep{RN1876}, where both CO\textsubscript{2} and H\textsubscript{2}O spectral features show strong evidence for scattering.

The actual HDO detection is the main outcome of this investigation. Still, it is interesting to ponder the question of why 132-1832 is so abundant in HDO ice if the quantitative result of a high HDO/H\textsubscript{2}O ratio is indeed valid. One possible explanation is isotopic H/D exchange in mixed ices of H\textsubscript{2}O/D\textsubscript{2}O, as demonstrated in laboratory experiments, starting at temperatures beyond $\sim$120~K and accelerating significantly at $\sim$150~K \citep{Galvez2011}; these are the conditions relevant to high temperature gradients in protoplanetary disks. Another possible explanation is more efficient thermal and photodesorption of H\textsubscript{2}O ice as compared to HDO ice. A higher HDO compared to H\textsubscript{2}O desorption energy was demonstrated in experiments of \citet{RN2015}. In addition, one of the conclusions of the photodesorption study of H\textsubscript{2}O, HDO, and D\textsubscript{2}O ices  by \citet{RN2013} was that photodesorption can lead to the enrichment of the ice in D atoms and, thus, to an enhanced HDO/H\textsubscript{2}O ratio. All these factors together may play a role in enhancing the HDO/H\textsubscript{2}O ratio. The increased photodesorption might be especially relevant for a radiatively elevated environment such as the location of 132-1832 in Orion, with the level of the UV intensity being of the order of several hundreds times the standard Habing unit $G_0$ \citep[cf.][]{2022RAA....22h5017X}. However, it is important to note that the HDO detection presented here is the first and, as of now, the only detection of HDO ice in protoplanetary disks. A dedicated survey covering sources in different radiation environments is required to draw firmer conclusions.

An open question remains as to what causes the dramatic drop of the HDO/H\textsubscript{2}O ratio during the transition from protoplanetary disks to minor bodies (comets, asteroids) of planetary systems. We note that we are referring here to the Solar System bodies shown in Figure~\ref{fig:ratio} and to rocky planets such as Earth. This question is out of the scope of the present study and requires further investigations. However, we would like to mention one of the possible explanations, which is linked to the phenomenon of strong bonding (trapping) of water molecules on silicates at temperatures above the desorption temperature of water ice \citep[inside the snowline;][]{RN990, RN1505, RN873, RN1715}. Our recent experimental results (Potapov et al., to be published) suggest that in mixed H$_2$O/D$_2$O ices H$_2$O molecules occupy stronger adsorption sites on silicate surfaces as compared to D\textsubscript{2}O molecules. Such an isotopolog-specific trapping may explain (at least partly) the lower D/H ratio on Earth and in primitive Solar System bodies as compared to the ISM and protoplanetary ices.

\section{Conclusions}
This study presents the first detection of HDO ice in protoplanetary disks. The estimated HDO/H\textsubscript{2}O ratio in the disk ices is higher than any other value previously determined for protostars and small bodies (i.e., comets, asteroids) in the Solar System. We consider the ratio as an upper limit for now. However, the result could potentially be explained by efficient isotopic H/D exchange in the disk ices and more efficient thermal and photodesorption of H\textsubscript{2}O compared to HDO ice. Thus, the observational results support laboratory experiments that simulate physico-chemical processes in deuterated cosmic ices. The ability of the hydrocarbon component to match the 3.4~\textmu m band opens up new avenues for probing the chemical and physical processing of ices and hydrocarbons in protoplanetary disks. Future JWST and ALMA observations will be crucial to determine whether the high HDO/H\textsubscript{2}O ratio observed in 132-1832 is a common characteristic of Class II disks as well as to further constrain the role of isotopic exchange and desorption processes. Extending such studies across larger disk samples will provide key insights into the chemical pathways that govern the delivery of water and organics to budding planetary systems.

\begin{acknowledgements}
We are grateful to Giulia Perotti for providing us Figure~4 and Tables~A5 and A6 and for fruitful discussions and to Katie Slavicinska for providing us the laboratory HDO data. This work is based on observations made with the NASA/ESA/CSA James Webb Space Telescope. The data were obtained from the Mikulski Archive for Space Telescopes at the Space Telescope Science Institute, which is operated by the Association of Universities for Research in Astronomy, Inc., under NASA contract NAS 5-03127 for JWST. The presented study was supported by the Federal Ministry for Economic Affairs and Climate Action on the basis of a decision by the German Bundestag (the German Aerospace Center project 50OR2215). AP and PK acknowledge support from the Deutsche Forschungsgemeinschaft (Heisenberg grant PO 1542/7-1 and research grant PO 1542/12-1). CB is grateful for an appointment at NASA Ames Research Center through the San Jos\'{e} State University Research Foundation (80NSSC22M0107) and acknowledges support from the Internal Scientist Funding Model (ISFM) Laboratory Astrophysics Directed Work Package Round 3 at NASA Ames.
   
\end{acknowledgements}

\bibliographystyle{aa} 
\bibliography{references}

\begin{appendix}





\onecolumn
\section{}
\begin{table*}[hb!]
\vspace{0.5em}
\centering
\caption{Properties of the disk 132-1832.}
\label{tab:disk_properties}
\begin{tabular}{lcl}
\hline\hline
Parameter & Value & Ref. \\
\hline
Spectral type & K5 & \citet{2012AJ....144..175T} \\
Distance & 390.2~pc & \citet{MaizApellaniz_2022} \\
Proj. Distance from OB stars & 294\textquotedbl19 & \citet{2012AJ....144..175T} \\
Disk mass & $0.80{\pm}0.08{\times}10^{-2}\ {\rm M}_\odot$ & \citet{2010ApJ...725..430M} \\
Projected disk size & 1\textquotedbl0 & \citet{2005AJ....129..382S} \\
Disk inclination & 75$^\circ$ & \citet{2000AJ....119.2919B} \\
\hline
\end{tabular}
\end{table*}

\begin{figure*}[ht]
\centering
\includegraphics[width=\hsize]{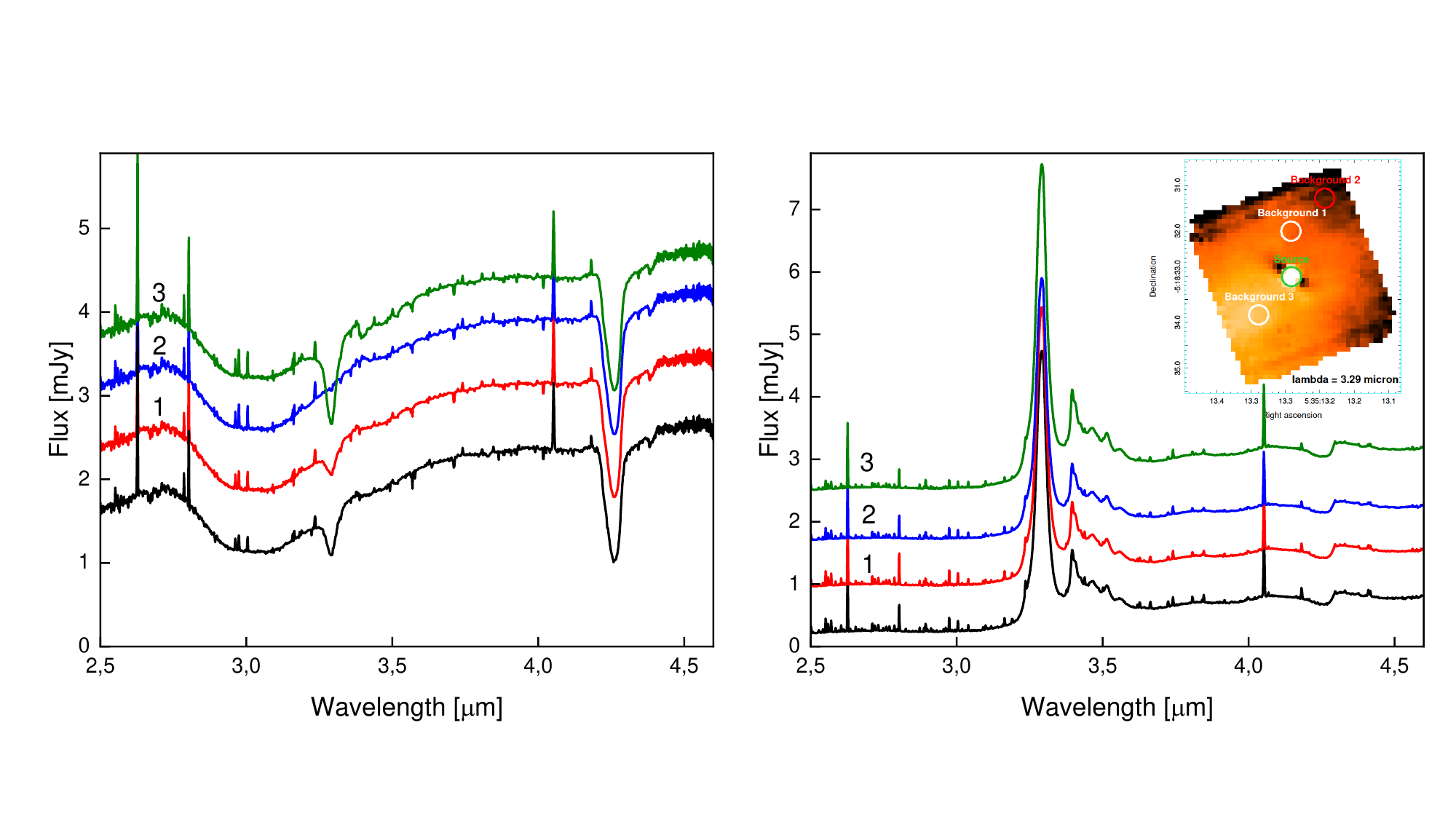}
\caption{Demonstrating the effect of the spatially varying PAH background when correcting the on-source spectrum. Left panel - the 2.5--4.6~\textmu m spectra, extracted at the source position and corrected by background estimates taken at the different positions indicated across the field of view. Right panel: Background spectra. The inset shows the brightness at 3.29~\textmu m, the spectral location of the PAH feature peak, and serves as a finding chart. The numbers (1, 2, 3) labeling the spectra correspond to the background positions shown in the inset. The three spectra are slightly offset vertically for clarity. The lower spectra are the standard data product utilising a sky annulus concentric background around the source (left) and the standard background (right).}
\label{fig:background}
\end{figure*}

In the resulting 1D spectra, a broad absorption feature is visible at the spectral location of the PAH band at around 3.29 micron. The occurrence of this feature is not an artifact of an incorrect background subtraction. It is a real feature that is independent
of the chosen background spectrum as shown in the figure below. The highly inclined circumstellar disk absorbs part of the extended PAH background emission from the Orion environment along its major axis. This leaves an absorption imprint in the background-subtracted source spectrum. Figure~\ref{fig:background} demonstrates the effect of the spatially varying PAH background.

The fitted HDO column density values based on spectra corrected with the customized backgrounds 1,2,3 are a bit lower than the value derived from the standard data product (utilising a sky annulus concentric around the source) but within the 50\% uncertainty limit mentioned in section 2.1: 4.6$\times$10$^{16}$~cm$^{\rm -2}$ (for Background 1), 3.4$\times$10$^{16}$~cm$^{\rm -2}$ (for Background 2) and 3.7$\times$10$^{16}$~cm$^{\rm -2}$ (for Background 3). The variance in the derived column densities depending on the chosen background may be attributed to the apparent gradient of the PAH and CO emission across the field of view and how it, after subtraction, influences the continuum determination. Notably, the continuum determination is susceptible to changes in the shape of the background broad band “PAH continuum” that starts around 3 µm with an initial jump and more-or-less plateaus after 3.5 µm as well as the CO rovibration structure around 4.7 µm. This emphasises that the classic evalution of an annulus around the source may be preferably, avoiding the complexity of non-local background emission and counteracting a simple background gradient to first degree. 

\begin{table}[ht!]
\caption{RMSE and AIC values obtained for various fits of the 2.7–4.0~\textmu m wavelength region. The parameters for the fit presented in Figure~\ref{fig:h20bestfit} are marked in \textbf{bold}.}
\label{tab:eniigmafit1}
\centering
\begin{tabular}{lcc}
\hline\hline
Input components&RMSE&AIC\\
\hline
Drude 1 + Drude 2 + MgSiO\textsubscript{3}/H\textsubscript{2}O 10K & 0.089 & 45.8 \\
Drude 1 + Drude 2 + MgSiO\textsubscript{3}/H\textsubscript{2}O 50K & 0.095 & 51.6 \\
Drude 1 + Drude 2 + MgSiO\textsubscript{3}/H\textsubscript{2}O 100K & 0.092 & 48.7 \\
Drude 1 + Drude 2 + MgSiO\textsubscript{3}/H\textsubscript{2}O 150K & 0.078 & 36.1 \\
\hline
Drude 1 + Drude 2 + MgSiO\textsubscript{3}/H\textsubscript{2}O/NH\textsubscript{3}* 10K & 0.061 & 24.5 \\
Drude 1 + Drude 2 + MgSiO\textsubscript{3}/H\textsubscript{2}O/NH\textsubscript{3} 50K & 0.059 & 23.3 \\
Drude 1 + Drude 2 + MgSiO\textsubscript{3}/H\textsubscript{2}O/NH\textsubscript{3} 100K & 0.062 & 25.3 \\
Drude 1 + Drude 2 + MgSiO\textsubscript{3}/H\textsubscript{2}O/NH\textsubscript{3} 150K & 0.083 & 40.9 \\
\hline
Drude 1 + Drude 2 + MgSiO\textsubscript{3}/H\textsubscript{2}O/CH\textsubscript{3}OH** 10K & 0.073 & 32.4 \\
Drude 1 + Drude 2 + MgSiO\textsubscript{3}/H\textsubscript{2}O/CH\textsubscript{3}OH 50K & 0.062 & 25.1 \\
Drude 1 + Drude 2 + MgSiO\textsubscript{3}/H\textsubscript{2}O/CH\textsubscript{3}OH 100K & 0.060 & 24.0 \\
Drude 1 + Drude 2 + MgSiO\textsubscript{3}/H\textsubscript{2}O/CH\textsubscript{3}OH 150K & 0.078 & 36.2 \\
\hline
Drude 1 + Drude 2 + MgSiO\textsubscript{3}/H\textsubscript{2}O 10K+50K & 0.091 & 49.5 \\
Drude 1 + Drude 2 + MgSiO\textsubscript{3}/H\textsubscript{2}O 10K+100K & 0.089 & 47.9 \\
Drude 1 + Drude 2 + MgSiO\textsubscript{3}/H\textsubscript{2}O 10K+150K & 0.079 & 39.6 \\
Drude 1 + Drude 2 + MgSiO\textsubscript{3}/H\textsubscript{2}O 50K+100K & 0.092 & 50.8 \\
Drude 1 + Drude 2 + MgSiO\textsubscript{3}/H\textsubscript{2}O 50K+150K & 0.079 & 38.9 \\
Drude 1 + Drude 2 + MgSiO\textsubscript{3}/H\textsubscript{2}O 100K+150K & 0.078 & 38.4 \\
\hline
Drude 1 + Drude 2 + MgSiO\textsubscript{3}/H\textsubscript{2}O/NH\textsubscript{3} 10K+50K & 0.059 & 25.5 \\
Drude 1 + Drude 2 + MgSiO\textsubscript{3}/H\textsubscript{2}O/NH\textsubscript{3} 10K+100K & 0.059 & 25.4 \\
Drude 1 + Drude 2 + MgSiO\textsubscript{3}/H\textsubscript{2}O/NH\textsubscript{3} 10K+150K & 0.063 & 28.0 \\
Drude 1 + Drude 2 + MgSiO\textsubscript{3}/H\textsubscript{2}O/NH\textsubscript{3} 50K+100K & 0.059 & 25.4 \\
Drude 1 + Drude 2 + MgSiO\textsubscript{3}/H\textsubscript{2}O/NH\textsubscript{3} 50K+150K & 0.057 & 25.6 \\
Drude 1 + Drude 2 + MgSiO\textsubscript{3}/H\textsubscript{2}O/NH\textsubscript{3} 100K+150K & 0.062 & 27.5 \\
\hline
Drude 1 + Drude 2 + MgSiO\textsubscript{3}/H\textsubscript{2}O/CH\textsubscript{3}OH 10K+50K & 0.063 & 28.2 \\
Drude 1 + Drude 2 + MgSiO\textsubscript{3}/H\textsubscript{2}O/CH\textsubscript{3}OH 10K+100K & 0.060 & 26.1 \\
Drude 1 + Drude 2 + MgSiO\textsubscript{3}/H\textsubscript{2}O/CH\textsubscript{3}OH 10K+150K & 0.062 & 27.6 \\
Drude 1 + Drude 2 + MgSiO\textsubscript{3}/H\textsubscript{2}O/CH\textsubscript{3}OH 50K+100K & 0.060 & 26.1 \\
Drude 1 + Drude 2 + MgSiO\textsubscript{3}/H\textsubscript{2}O/CH\textsubscript{3}OH 50K+150K & 0.061 & 26.5 \\
Drude 1 + Drude 2 + MgSiO\textsubscript{3}/H\textsubscript{2}O/CH\textsubscript{3}OH 100K+150K & 0.061 & 26.7 \\
\hline
Drude 1 + Drude 2 + MgSiO\textsubscript{3}/H\textsubscript{2}O/NH\textsubscript{3} 10K+Chemistry*** & 0.028 & 12.1 \\
Drude 1 + Drude 2 + MgSiO\textsubscript{3}/H\textsubscript{2}O/NH\textsubscript{3} 50K+Chemistry & 0.031 & 13.0 \\
Drude 1 + Drude 2 + MgSiO\textsubscript{3}/H\textsubscript{2}O/NH\textsubscript{3} 100K+Chemistry & 0.040 & 16.3 \\
Drude 1 + Drude 2 + MgSiO\textsubscript{3}/H\textsubscript{2}O/NH\textsubscript{3} 150K+Chemistry & 0.063 & 27.9 \\
Drude 1 + Drude 2 + MgSiO\textsubscript{3}/H\textsubscript{2}O/CH\textsubscript{3}OH 10K+Chemistry & 0.029 & 12.2 \\
Drude 1 + Drude 2 + MgSiO\textsubscript{3}/H\textsubscript{2}O/CH\textsubscript{3}OH 50K+Chemistry & 0.027 & 11.7 \\
Drude 1 + Drude 2 + MgSiO\textsubscript{3}/H\textsubscript{2}O/CH\textsubscript{3}OH 100K+Chemistry & 0.033 & 13.6 \\
Drude 1 + Drude 2 + MgSiO\textsubscript{3}/H\textsubscript{2}O/CH\textsubscript{3}OH 150K+Chemistry & 0.060 & 26.2 \\
\hline
\textbf{Drude 1 + Drude 2 + MgSiO\textsubscript{3}/H\textsubscript{2}O/CH\textsubscript{3}OH 50K+Chemistry+Hydrocarbons****} & \textbf{0.022} & \textbf{12.7} \\
\hline
\end{tabular}
\tablefoot{* H\textsubscript{2}O/NH\textsubscript{3} ratio is 12:1. ** H\textsubscript{2}O/CH\textsubscript{3}OH ratio is 10:1. *** ``Chemistry'' refers to the laboratory spectrum obtained after UV irradiation of a mixture of CO\textsubscript{2}+NH\textsubscript{3} at 75~K and its subsequent heating to 230~K \citep{RN983}, mainly containing ammonium carbamate and carbamic acid, used for fitting the 3~\textmu m band in \citet{RN1876}. **** ``Hydrocarbons'' refers to the laboratory spectrum obtained after UV irradiation of H\textsubscript{2}O+CH\textsubscript{3}OH+MgSiO\textsubscript{3} at 10~K and subsequent heating to 200~K, used for fitting the 3~\textmu m band in \citep{RN873}.}
\end{table}

\begin{table}[ht!]
\caption{RMSE and AIC values obtained for various fits of the 4.0–4.45~\textmu m wavelength region. The parameters for the fit presented in Figure~\ref{fig:hdofit} are marked in \textbf{bold}.}
\label{tab:eniigmafit2}
\centering
\begin{tabular}{lcc}
\hline\hline
Input components&RMSE&AIC\\
\hline
CO\textsubscript{2} + HDO08* 14K& 0.057 & 20.2 \\
CO\textsubscript{2} + HDO4* 14K& 0.056 & 19.6 \\
CO\textsubscript{2} + HDO20* 14K & 0.056 & 19.6 \\
CO\textsubscript{2} + HDO20 90K & 0.056 & 19.7 \\
CO\textsubscript{2} + HDO4 90K  & 0.057 & 20.2 \\
CO\textsubscript{2} + HDO20 150K & 0.055 & 19.4 \\
CO\textsubscript{2} + HDO4 150K  & 0.056 & 19.5 \\
\hline
H\textsubscript{2}O/CO\textsubscript{2} + HDO08 14K& 0.122 & 79.2 \\
H\textsubscript{2}O/CO\textsubscript{2} + HDO4 14K& 0.119 & 74.4 \\
H\textsubscript{2}O/CO\textsubscript{2} + HDO20 14K& 0.114 & 69.2 \\
H\textsubscript{2}O/CO\textsubscript{2} + HDO20 90K& 0.114 & 68.5 \\
H\textsubscript{2}O/CO\textsubscript{2} + HDO4 90K & 0.123 & 79.2 \\
H\textsubscript{2}O/CO\textsubscript{2} + HDO20 150K & 0.115 & 70.2 \\
H\textsubscript{2}O/CO\textsubscript{2} + HDO4 150K  & 0.120 & 76.0 \\
\hline
CO\textsubscript{2} + HDO20 14K + H\textsubscript{2}O/CO\textsubscript{2} & 0.050 & 18.3 \\
CO\textsubscript{2} + HDO20 14K + CO\textsubscript{2}\ scattering 13K & 0.049 & 18.0 \\
\textbf{CO\textsubscript{2} + HDO20 14K + CO\textsubscript{2}/CO }& \textbf{0.046} &\textbf{16.6}\\
\hline
CO\textsubscript{2} + HDO20 (14K + 150K) + CO\textsubscript{2}/CO & 0.046 & 18.5 \\
CO\textsubscript{2} + HDO20 14K + CO\textsubscript{2}/CO + H\textsubscript{2}O/CO\textsubscript{2} & 0.046 & 18.8 \\
CO\textsubscript{2} + HDO20 14K + CO\textsubscript{2}/CO + CO\textsubscript{2} scattering 13K & 0.046 & 18.5 \\
CO\textsubscript{2} + HDO20 14K + CO\textsubscript{2}/CO + CH\textsubscript{3}OH
& 0.046 
& 18.5 \\
\hline
CO\textsubscript{2} + HDO(c)\textsuperscript{***} 15K + CO\textsubscript{2}/CO & 0.047 & 16.9 \\
CO\textsubscript{2} + HDO(c) 25K + CO\textsubscript{2}/CO & 0.047 & 16.8 \\
CO\textsubscript{2} + HDO(c) 68K + CO\textsubscript{2}/CO & 0.047 & 16.9 \\
CO\textsubscript{2} + HDO(c) 111K + CO\textsubscript{2}/CO & 0.047 & 16.9 \\
CO\textsubscript{2} + HDO(c) 133K + CO\textsubscript{2}/CO & 0.047 & 17.0 \\
CO\textsubscript{2} + HDO(c) 150K + CO\textsubscript{2}/CO & 0.047 & 17.0 \\
\hline
CO\textsubscript{2} + HDO(a)\textsuperscript{***} 15K + CO\textsubscript{2}/CO & 0.048 & 17.4 \\
CO\textsubscript{2} + HDO(a) 31K + CO\textsubscript{2}/CO & 0.048 & 17.3 \\
CO\textsubscript{2} + HDO(a) 51K + CO\textsubscript{2}/CO & 0.048 & 17.2 \\
CO\textsubscript{2} + HDO(a) 71K + CO\textsubscript{2}/CO & 0.048 & 17.1 \\
CO\textsubscript{2} + HDO(a) 90K + CO\textsubscript{2}/CO & 0.048 & 17.1 \\
CO\textsubscript{2} + HDO(a) 111K + CO\textsubscript{2}/CO & 0.047 & 17.1 \\
CO\textsubscript{2} + HDO(a) 131K + CO\textsubscript{2}/CO & 0.047 & 17.0 \\
CO\textsubscript{2} + HDO(a) 141K + CO\textsubscript{2}/CO & 0.047 & 17.0 \\
\hline
\end{tabular}
\tablefoot{*08, 4 and 20 indicate 0.8\%, 4\%, and 20\% of HDO in H\textsubscript{2}O in the laboratory sample,. For details, see \citet{Galvez2011}.**Denotes CO\textsubscript{2} scattering at different temperatures. For details, see Section 3.2 and Table 3 in \citet{RN1876}.
***HDO(c) denotes crystalline HDO, and HDO(a) denotes amorphous HDO. For details, see \citet{Slavicinska2024}.}
\end{table}

\begin{figure*}[ht]
\centering

\begin{minipage}{0.31\textwidth}
\centering
\includegraphics[width=\textwidth]{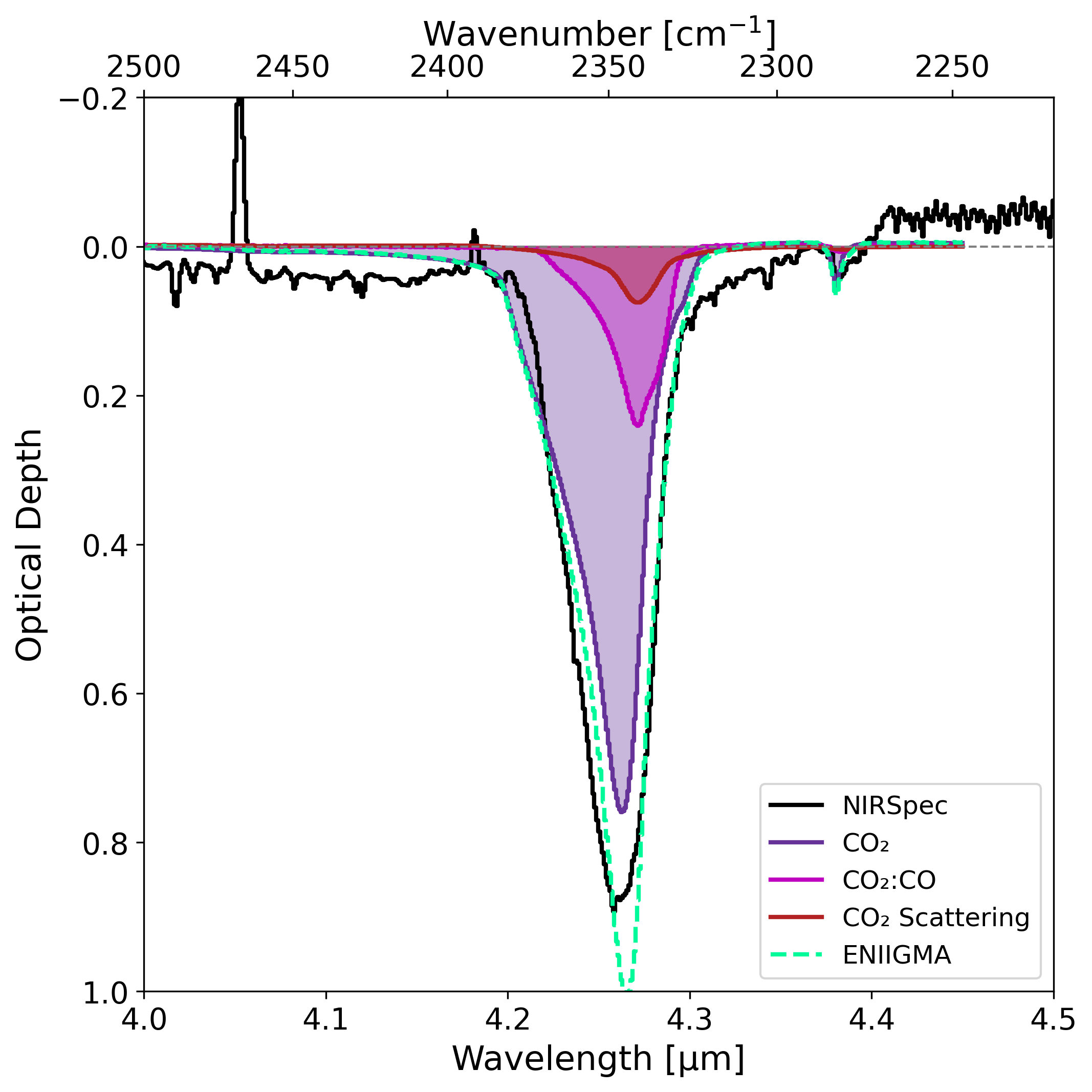}
\par\vspace{0.1cm}
(A)
\end{minipage}
\hfill
\begin{minipage}{0.31\textwidth}
\centering
\includegraphics[width=\textwidth]{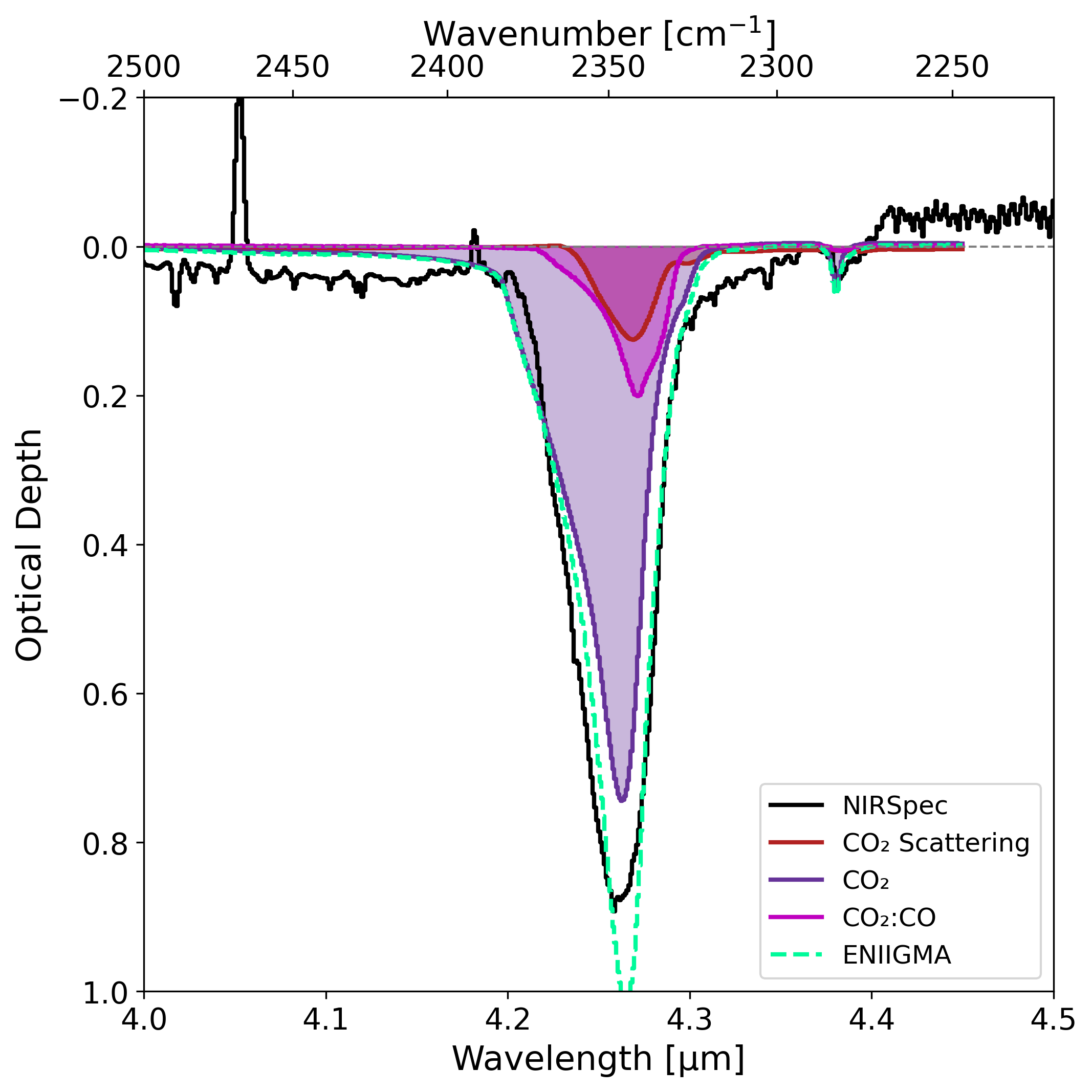}
\par\vspace{0.1cm}
(B)
\end{minipage}
\hfill
\begin{minipage}{0.31\textwidth}
\centering
\includegraphics[width=\textwidth]{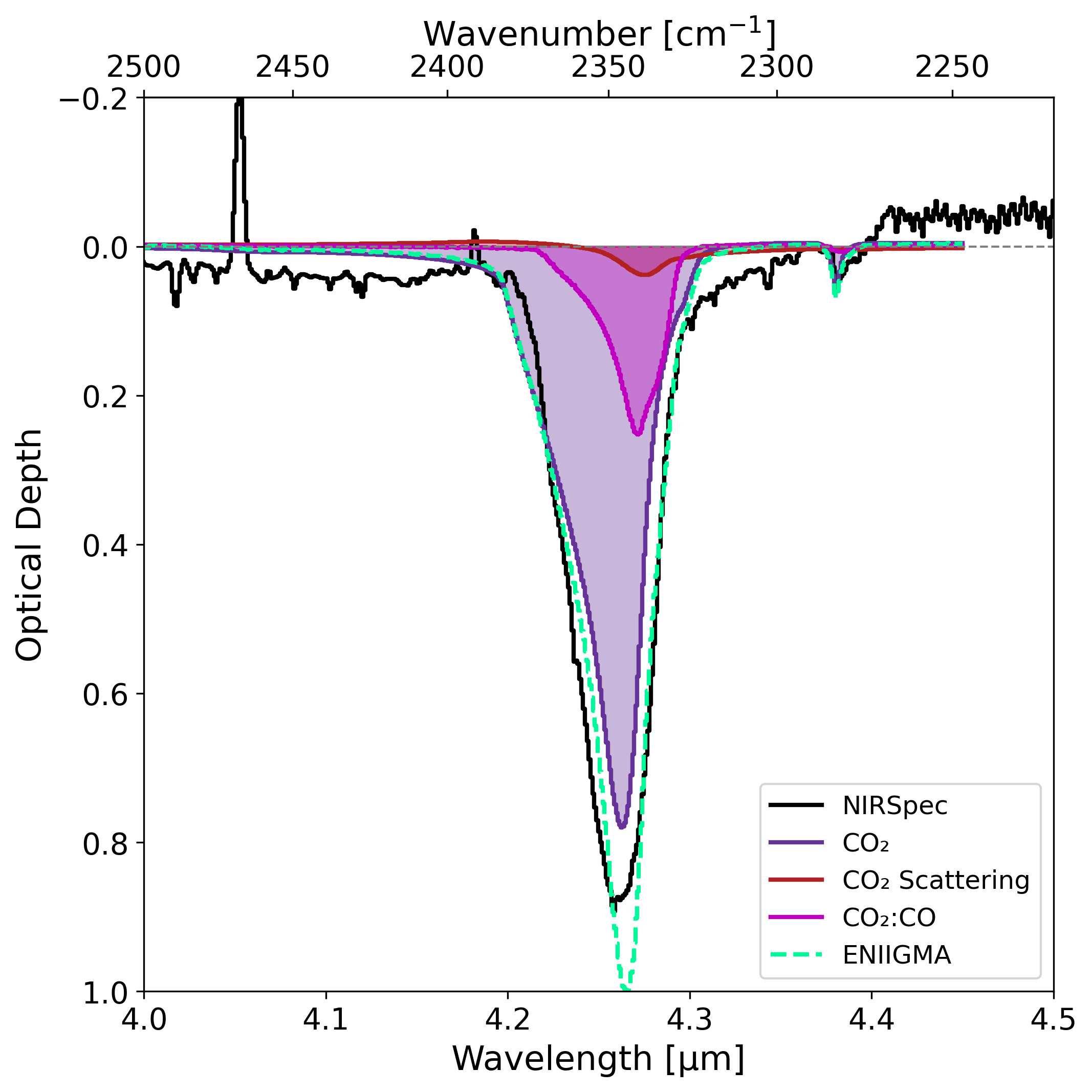}
\par\vspace{0.1cm}
(C)
\end{minipage}

\vspace{0.35cm}

\begin{minipage}{0.31\textwidth}
\centering
\includegraphics[width=\textwidth]{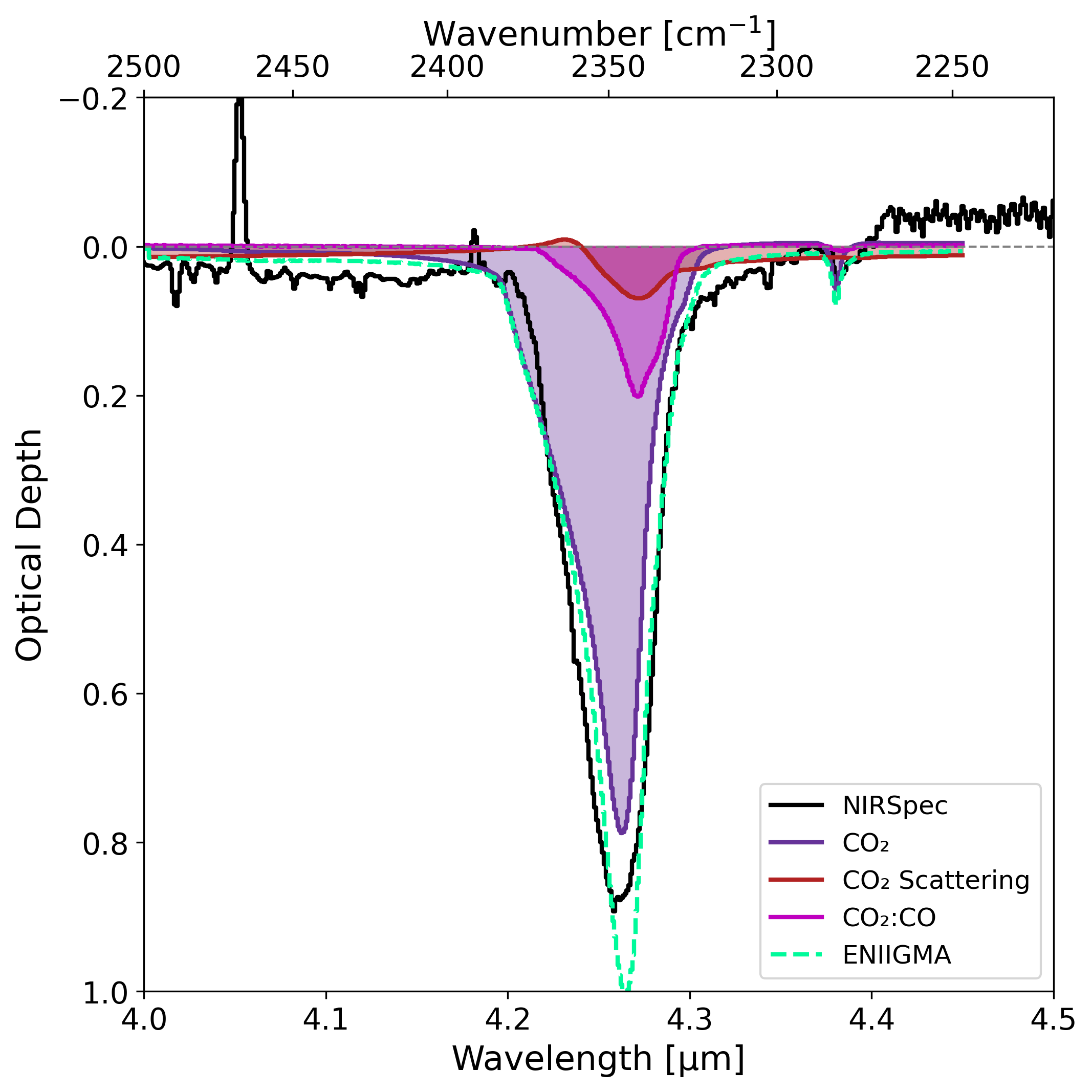}
\par\vspace{0.1cm}
(D)
\end{minipage}
\hfill
\begin{minipage}{0.31\textwidth}
\centering
\includegraphics[width=\textwidth]{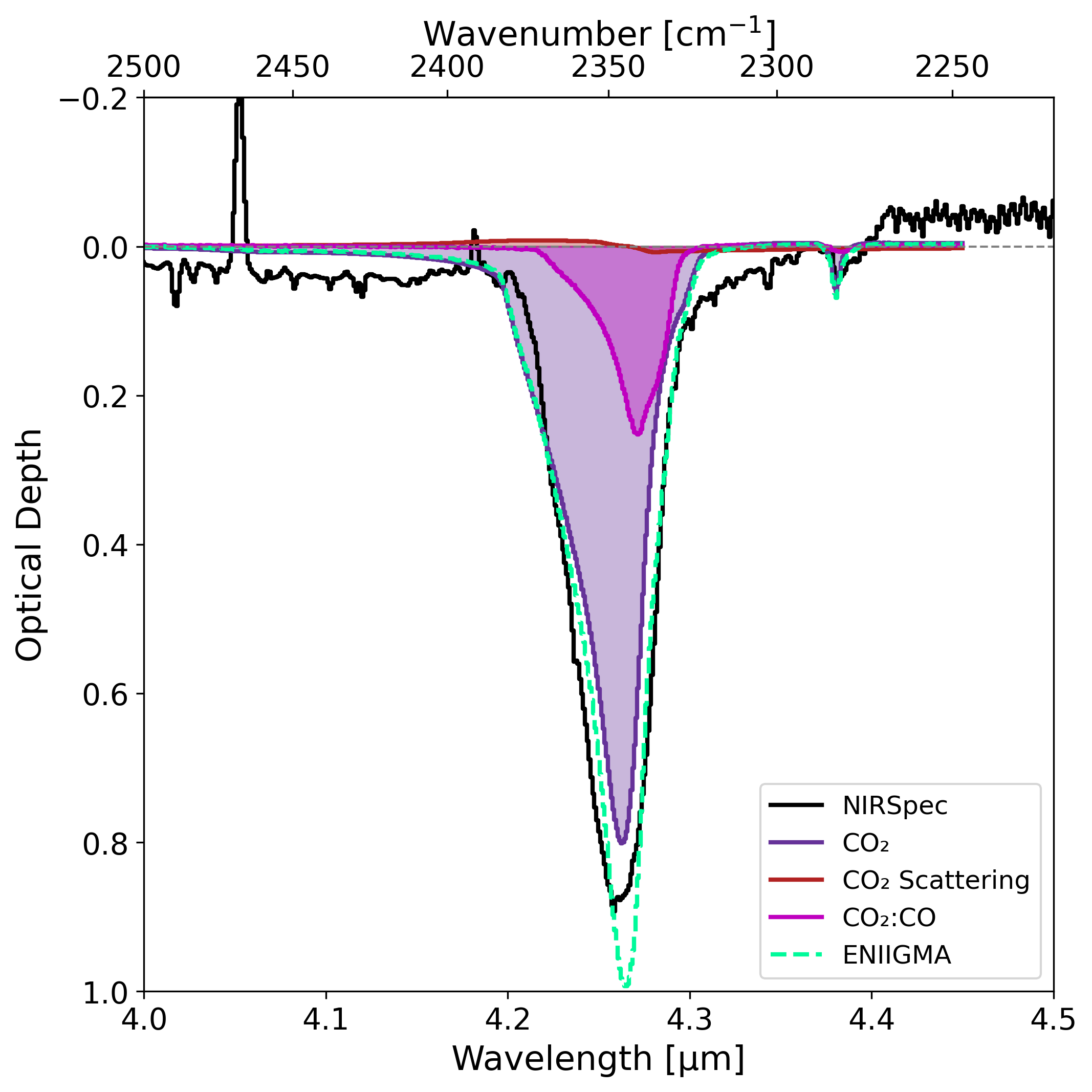}
\par\vspace{0.1cm}
(E)
\end{minipage}
\hfill
\begin{minipage}{0.31\textwidth}
\centering
\includegraphics[width=\textwidth]{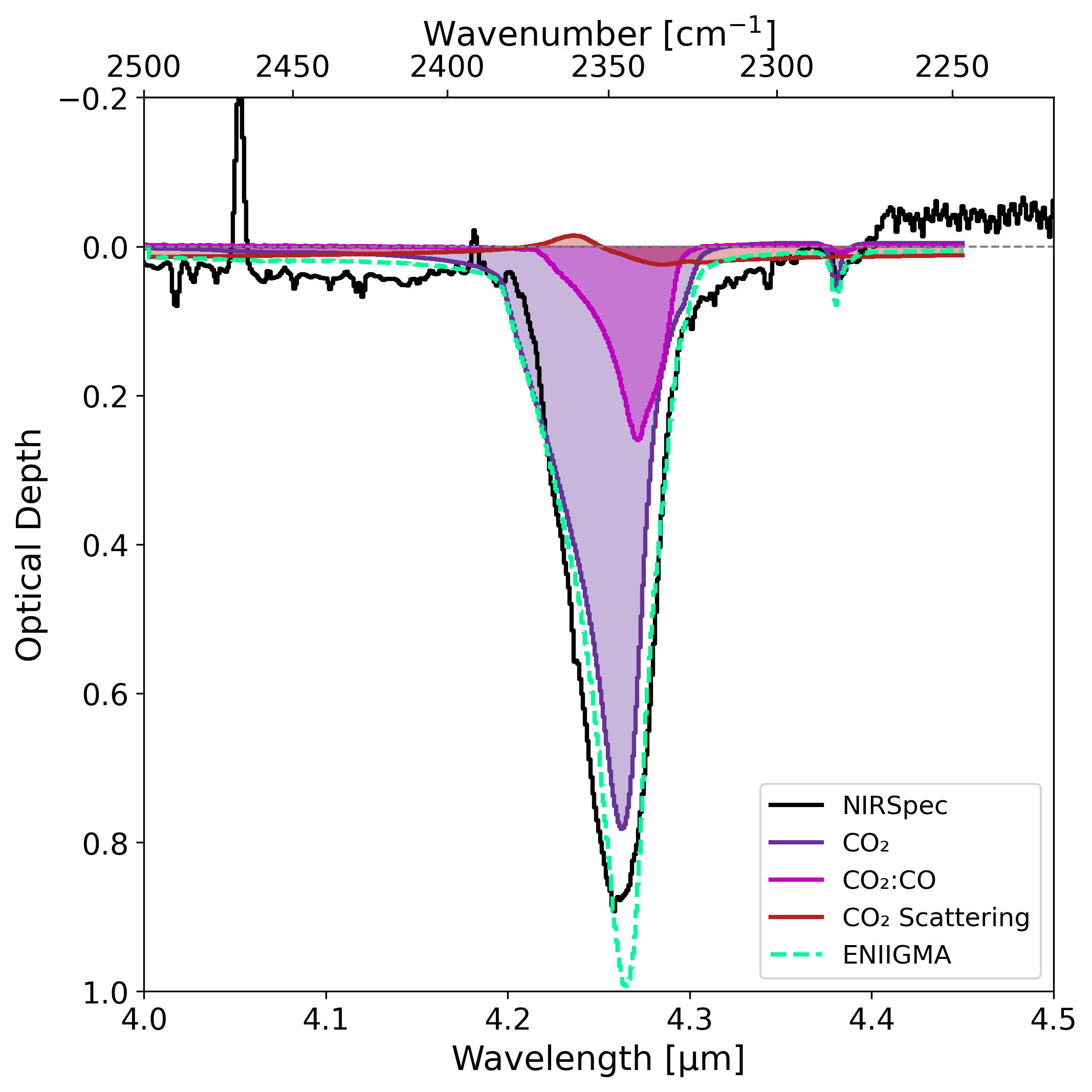}
\par\vspace{0.1cm}
(F)
\end{minipage}

\caption{Optical depth spectrum and ENIIGMA fits for the 4.0--4.45~\textmu m region including CO\textsubscript{2} scattering models computed with \texttt{OpTool}. Panels A--F correspond to the model sets summarized in Table~\ref{tab:scattering}.}
\label{fig:co2_scattering_AF}
\end{figure*}

Figure~\ref{fig:co2_scattering_AF} shows the fit models for the edge cases (a$_m$$_a$$_x$ = 1 and 5~\textmu m and the power law indices of -3.5 and -2.0).

\begin{table}
\caption{Summary of the CO\textsubscript{2} scattering models shown in Figure~\ref{fig:co2_scattering_AF}.}
\label{tab:scattering}
\centering
\begin{tabular}{c c c c}
\hline\hline
Model & Ice mantle & $a_{\max}$ (\textmu m) & Power-law index \\
\hline
A & CO\textsubscript{2} & 1.0 & $-3.5$ \\
B & H\textsubscript{2}O:CO\textsubscript{2} (10:1) & 1.0 & $-3.5$ \\
C & CO\textsubscript{2} & 5.0 & $-3.5$ \\
D & H\textsubscript{2}O:CO\textsubscript{2} (10:1) & 5.0 & $-3.5$ \\
E & CO\textsubscript{2} & 5.0 & $-2.0$ \\
F & H\textsubscript{2}O:CO\textsubscript{2} (10:1) & 5.0 & $-2.0$ \\
\hline
\end{tabular}
\tablefoot{All models use $a_{\min}=0.005~\mu$m and the same MgSiO\textsubscript{3} dust core.}
\end{table}

\begin{table}[ht!]
\caption{Ice and gas HDO/H\textsubscript{2}O values of Solar System objects. The values are graphically shown in Figure~\ref{fig:ratio}.}
\label{tab:C3}
\centering
\begin{tabular}{llcc}
\hline\hline
Object&Phase&HDO/H\textsubscript{2}O ($10^{-3}$)&Ref\\
\hline
\multicolumn{4}{c}{\textit{Carbonaceous chondrites}} \\
\hline
CI chondrites     & — & $0.129$--$0.195$ & 1\\
CM chondrites     & — & $0.173 \pm 0.007$ & 1\\
CR chondrites     & — & $0.342$ & 1\\
CO chondrites     & — & $0.170$--$0.264$ & 1\\
\hline
\multicolumn{4}{c}{\textit{Halley-type comets}} \\
\hline
12P/Pons-Brooks     & gas & $0.17 \pm 0.04$ & 2\\
1P/Halley           & gas & $0.42 \pm 0.06$ & 3\\
8P/Tuttle           & gas & $0.82 \pm 0.29$ & 4\\
153P/Ikeya-Zhang    & gas & $<0.56 \pm 0.06$ & 5\\
\hline
\multicolumn{4}{c}{\textit{Jupiter-family comets}} \\
\hline
46P/Wirtanen        & gas & $0.32 \pm 0.13$ & 6\\
103P/Hartley 2      & gas & $0.32 \pm 0.05$ & 7\\
67P/Churyumov–Gerasimenko & gas & $1.06 \pm 0.14$ & 8\\
45P/Honda-Mrkos-Pajdušáková & gas & $<0.40$ & 9\\
\hline
\multicolumn{4}{c}{\textit{Oort-cloud comets}} \\
\hline
C/2009 P1 Garradd   & gas & $0.41 \pm 0.04$ & 10\\
C/2002 T7 LINEAR    & gas & $0.50 \pm 0.14$ & 11\\
C/1996 B2 Hyakutake & gas & $0.58 \pm 0.20$ & 12\\
C/2014 Q2 Lovejoy   & gas & $0.28 \pm 0.08$ & 13\\
C/2012 F6 Lemmon    & gas & $1.30 \pm 0.32$ & 13\\
C/1995 O1 Hale-Bopp & gas & $0.66 \pm 0.16$ & 14\\
C/2007 B3 Lulin     & gas & $<1.12$ &  15\\
\hline
\end{tabular}
\tablefoot{Refs. 1. \citet{2012Sci...337..721A} 2. \citet{2025NatAs.tmp..159C} 3. \citet{2012P&SS...60..166B} 4. \citet{2009ApJ...690L...5V} 5. \citet{2006A&A...449.1255B} 6. \citet{2019A&A...625L...5L} 7. \citet{2011Natur.478..218H} 8. \citet{Altwegg2015} 9. \citet{2013ApJ...774L...3L} 10. \citet{2012A&A...544L..15B} 11. \citet{2008A&A...490L..31H} 12. \citet{1998Icar..133..147B} 13. \citet{2016A&A...589A..78B} 14. \citet{1998Sci...279..842M} 15. \citet{2012ApJ...750..102G}}
\end{table}

\begin{table}[ht!]
\caption{Ice and gas HDO/H\textsubscript{2}O values of protostars. The values are graphically shown in Figure~\ref{fig:ratio}}
\label{tab:C4}
\centering
\begin{tabular}{llcc}
\hline\hline
Object&Phase&HDO/H\textsubscript{2}O ($10^{-3}$)&Ref\\
\hline
\multicolumn{4}{c}{\textit{Clustered class 0 LYSOs}} \\
\hline
NGC 1333 IRAS 4A-NW & gas & $0.54 \pm 0.15$ &1, 2\\
NGC 1333 IRAS 2A & gas & $0.74 \pm 0.21$ &1\\
NGC 1333 IRAS 4B & gas & $0.59 \pm 0.26$ &1\\
IRAS 16293-2422 & gas & $0.92 \pm 0.26$ &1\\
NGC 1333 SVS 13 & ice & $\leq 17$ &3\\
\hline
\multicolumn{4}{c}{\textit{Clustered class I LYSOs}} \\
\hline
NGC 1333 SVS 12 & ice & $\leq 5$ &3\\
\hline
\multicolumn{4}{c}{\textit{Isolated class 0 LYSOs}} \\
\hline
BHR 71-IRS1 & gas & $1.8 \pm 0.4$ &2 \\
B335 & gas & $6.3 \pm 1.5$ &4\\
L483 & gas & $4.0 \pm 0.5$ &4\\
\hline
\multicolumn{4}{c}{\textit{Isolated class 0/I LYSOs}} \\
\hline
L1527 IRS & ice & $4.4^{+3.7}_{-1.7}$ &5\\
\hline
\multicolumn{3}{c}{\textit{Isolated class I LYSOs}} \\
\hline
V883 Ori & gas & $2.3 \pm 0.6$ &6\\
L1551 IRS5 & gas & $2.1 \pm 0.8$ &7\\
L1489 IRS & ice & $\leq 8$ &3\\
TMR1 & ice & $\leq 11$ &3\\
\hline
\multicolumn{4}{l}{\textit{Class II LYSOs}} \\
\hline
132-1832 & ice & $\leq 51$ & This work\\
\hline
\multicolumn{4}{l}{\textit{Intermediate-mass YSOs (IMYSOs)}} \\
\hline
IRAS 05390-0728 & ice & $\leq 10$ & 8\\
IRAS 08448-4343 & ice & $\leq 10$ & 8\\
HOPS 370 & ice & $4.6 \pm 2.2$ & 9\\
\hline
\multicolumn{4}{c}{\textit{Massive YSOs (MYSOs)}} \\
\hline
W3 IRS5 & gas & $1.3$ & 10\\
W33A & gas & $3.0$ & 10\\
AFGL 2591 & gas & $3.3$ &10\\
NGC 7538 IRS1 & gas & $3.8$ &10\\
Orion KL Hot Core & gas & $3.0^{+3.1}_{-1.7}$ &11\\
NGC 6334 I & gas & $0.21 \pm 0.10$ &12\\
G34.26+0.15 & gas & $0.35$--$0.75$ &13\\
NGC 7538 IRS9 & ice & $\leq 8.1$--$11.4$& 8\\
GL 2136 & ice & $\leq 4$ &8\\
IRAS 20126 & ice & $2.6 \pm 1.4$ &9\\
\hline
\end{tabular}
\tablefoot{Refs. 1. \citet{Persson2014} 2. \citet{Jensen2019} 3. \citet{2003A&A...410..897P} 4. \citet{Jensen2021b} 5. \citet{Slavicinska2025} 6. \citet{Tobin2023} 7. \citet{Andreu2023} 8. \citet{2003A&A...399.1009D} 9. \citet{Slavicinska2024} 10. \citet{2006A&A...447.1011V} 11. \citet{2013ApJ...777...85N} 12. \citet{2013ApJ...765...61E} 13. \citet{2014MNRAS.445.1299C}}
\end{table}

\FloatBarrier 
\clearpage
\end{appendix}
\end{document}